\documentclass[preprint, authoryear]{elsarticle}
\usepackage{amsmath,amsfonts,amssymb}
\usepackage{graphicx}
\usepackage{epsf}
\usepackage{here}
\usepackage{graphicx}
\usepackage[american]{babel}

\begin{document}

\title{Temporal structure and gain/loss asymmetry for real and artificial stock indices}

\author[saxo]{Johannes Vitalis Siven}
\ead{jvs@saxobank.com}

\author[saxo]{Jeffrey Todd Lins}
\ead{jtl@saxobank.com}

\address[saxo]{Saxo Bank A/S, Philip Heymans All\'e 15, DK-2900 Hellerup, Denmark}

\begin{abstract} We demonstrate that the gain/loss asymmetry observed
for stock indices vanishes if the temporal dependence structure is
destroyed by scrambling the time series. We also show that an
artificial index constructed by a simple average of a number of
individual stocks display gain/loss asymmetry --- this allows us
to explicitly analyze the dependence between the index
constituents. We consider mutual information and correlation based
measures and show that the stock returns indeed have a higher
degree of dependence in times of market downturns than upturns.
\end{abstract}


\maketitle

\section*{Introduction} 
\begin{figure}
\begin{center}
\includegraphics[width=12cm]{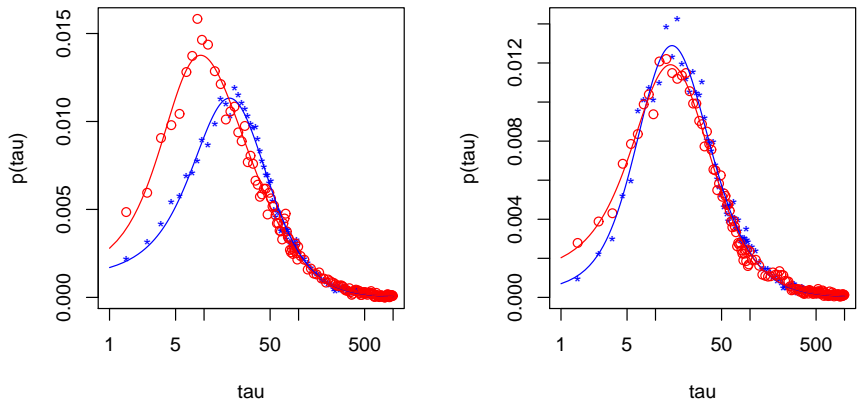}
\end{center}
\caption{Estimated distribution of the first passage time
$\tau_\rho$ for the log price of the Dow Jones Industrial Average
index (left) and its scrambled version (right). The graphs
correspond to $\rho = +5\%$ (stars) and $\rho = -5\%$ (rings). The
solid lines are fitted generalized gamma density functions.}
\label{fig:DJIA_FPT}
\end{figure}

Inspired by research in the field of turbulence,
Simonsen, Jensen, and Johansen \cite{investHorizon} considered
``inverse statistics'' of financial time series: what is the
smallest time interval needed for an asset to cross a fixed return
level $\rho$? Figure \ref{fig:DJIA_FPT} shows the distribution of
this random variable, the \emph{first passage time}, for the Dow
Jones Industrial Average index, for $\rho = \pm 5\%$. As noted by
Jensen, Johansen, and Simonsen \cite{investStatistics}, the most
likely first passage time is shorter for $\rho = -5\%$ than for
$\rho = 5\%$, which they refer to as the \emph{gain/loss
asymmetry}.

In this paper, we show that the gain/loss asymmetry in the Dow
Jones index vanishes if the time series is ``scrambled''
--- that is, if one considers a new time series constructed by
randomly permuting the returns. This basic fact, which seems to
have gone unnoticed in the literature so far, has important
implications: the gain/loss asymmetry is \emph{not} due to
properties of the unconditional index returns, like skewness, but
is rather an expression of potentially complex temporal structure.
This finding resonates with the results from Siven, Lins, and
Lundbek~Hansen \cite{multiscale}, where wavelet analysis is used
to demonstrate that the gain/loss asymmetry is a long time scale
phenomenon
--- it vanishes if enough low frequency content is removed from
the index, that is, if the index is sufficiently ``detrended.''
Siven, Lins, and Lundbek~Hansen \cite{multiscale} also present a
generalization of the asymmetric synchronous market model from
Donangelo, Jensen, Simonsen, and Sneppen \cite{fearModel} where
prolonged periods of high correlation between the individual
stocks during index downturns gives rise to a gain/loss asymmetry.

Whether the constituents of e.g.\ the Dow Jones index indeed tend
to move with a greater degree of dependence during market
downturns could in principle be tested empirically by analysis of
the time series of the individual stocks. That is an awkward task,
however, since the relative weights for different stocks in these
indices have changed over time in complicated ways. To address
this issue, we demonstrate that if one defines a new, artificial
index by simply taking the average of a number of stocks, this
index also displays gain/loss asymmetry. With the constituents
readily available, we consider two measures based on correlation
and mutual information, and show that there indeed is a higher
degree of dependence between the stock returns during index
downturns than upturns.


\section*{Gain/loss asymmetry and temporal structure}
For a given process $\{I_t\}_{t \geq 0}$, for instance daily
closing prices of a stock index, the first passage time
$\tau_\rho$ of the level $\rho$ is defined as
$$
\tau_\rho = \left\{ \begin{array}{ll}
\inf\{s>0;\ \log(I_{t+s}/I_t) \geq \rho\}& \mbox{if } \rho> 0,\\
 \inf\{s>0;\ \log(I_{t+s}/I_t) \leq \rho\}& \mbox{if } \rho< 0,
\end{array} \right.
$$
and is assumed to be independent of $t$. The distribution of
$\tau_\rho$ is estimated in a straightforward manner from a time
series $I_{0},\ldots,I_{T}$. Consider $\rho > 0$, and let $t+s$ be
the smallest time point such that $\log(I_{t+s}/I_{t}) \geq \rho$,
if such a time point exists. In that case, $s$ is viewed as an
observation of $\tau_\rho$. (If $\rho <0$, take instead $t+s$ such
that $\log(I_{t+s}/I_{t}) \leq \rho$.) Running $t$ from $0$ to
$T-1$ gives a set of observations from which the distribution of
$\tau_\rho$ is estimated as the empirical distribution. Given the
empirical distribution, we follow Jensen, Johansen, and Simonsen
\cite{investStatistics} and compute a fit of the density function
for the generalized gamma distribution. This density is plotted as
a solid line together with the empirical distribution in all
figures, to guide the eye
--- we do not discuss the fitted parameters, nor claim that
$\tau_\rho$ truly follows a generalized gamma distribution.

\subsection*{Gain/loss asymmetry for the Dow Jones index}
Figure \ref{fig:DJIA_FPT} shows the estimated first passage time
for the Dow Jones index. As discussed in the Introduction, there
is a gain/loss asymmetry in that the most likely first passage
time is shorter for $\rho = -5\%$ than for $\rho = 5\%$. Next, we
construct a \emph{scrambled} version of the index by randomly
re-arranging the log returns. Formally, if $I_0,\ldots,I_T$
denotes the time series of daily closing prices of the Dow Jones
index, let $\delta I_{t} = \log (I_t /I_{t-1})$ for $t =
1,\ldots,T$ and draw a random permutation $j_1,\ldots,j_T$ of
$\{1,\ldots,T\}$. We define
$$
\delta \tilde{I}_t = \delta I_{j_t}, \qquad \textrm{for }t =
1,\ldots,T,
$$
and let the scrambled index be given by $\tilde{I}_0 = I_0$ and
$$
\tilde{I}_t = \tilde{I}_0\exp\left(\sum_{s = 1}^t \delta
\tilde{I}_s\right), \qquad \textrm{for }t = 1,\ldots,T.
$$
Figure \ref{fig:DJIA_FPT} shows that the scrambled index does
\emph{not} display a gain/loss asymmetry. This result is
surprisingly strong: since the empirical return distributions are
identical for an index and any of its scrambled versions, it shows
that the gain/loss asymmetry is an expression of potentially
complex temporal structure in the index. This fits nicely with the
results from Siven, Lins, and Lundbek~Hansen \cite{multiscale},
where a multiscale decomposition is used to demonstrate that the
gain/loss asymmetry is a long rather than short scale phenomenon.

\begin{figure}
\begin{center}
\includegraphics[width=12cm]{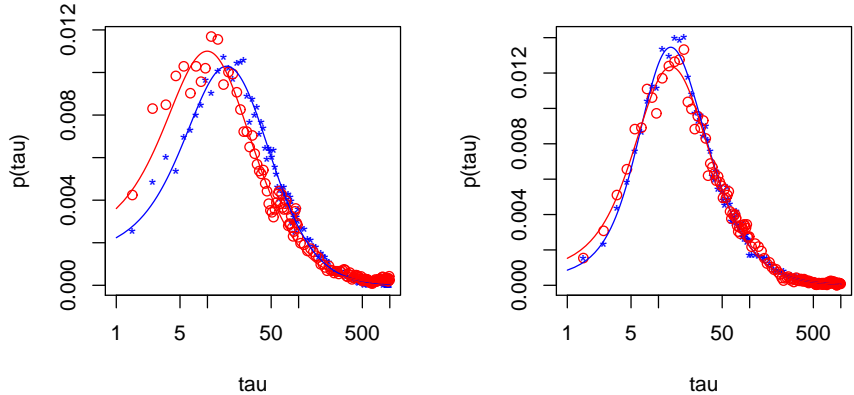}
\end{center}
\caption{Estimated distribution of the first passage time
$\tau_\rho$ for the log price in a realization of the generalized
asynchronous market model from Siven, Lins, and Lundbek~Hansen
\cite{multiscale} (left) and its scrambled version (right). The
graphs correspond to $\rho = +5\%$ (stars) and $\rho = -5\%$
(rings). The solid lines are fitted generalized gamma density
functions.} \label{fig:GASMM}
\end{figure}

The gain/loss asymmetry in the asymmetric synchronous market model
from Donangelo, Jensen, Simonsen, and Sneppen \cite{fearModel}
does not disappear when the index returns are scrambled. This is
to be expected, since the daily returns in that model are
independent and identically distributed, so all statistical
properties remain the same when the time series is scrambled.
However, for the generalized model proposed in Siven, Lins, and
Lundbek~Hansen \cite{multiscale} the asymmetry \emph{does} vanish,
in perfect agreement with the Dow Jones index, see Figure
\ref{fig:GASMM}.

\subsection*{Gain/loss asymmetry for an artificial index} Consider $N$ stocks, and let $S_{n,t}$ denote the
closing price of the $n$th stock on day $t$, for $t =
0,1,\ldots,T$. We consider the artificial index constructed by
averaging all the stocks,
$$I_t = \frac{1}{N}\sum_{n =
1}^N\frac{S_{n,t}}{S_{n,0}}.
$$
The denominators $S_{n,0}$ give all stocks equal weight in the
index at time $t = 0$.

We consider historical stock prices from January 1970 until
December 2008 for the following 12 Dow Jones constituents: Boeing
Co.~(BA), Citigroup Inc.~(C), El DuPont de Nemours \& Co.~(DD),
General Electric Co.~(GE), General Motors Corporation (GM),
International business Machines Corp.~(IBM), Johnson \& Johnson
(JNJ), JPMorgan Chase \& Co.~(JPM), The Coca-Cola Company (KO),
McDonald's Corp.~(MCD), Procter \& Gamble Co.~(PG), and Alcoa
Inc.~(AA). These companies are chosen since long time series of
stock returns are available, but our results are stable in the
sense that adding or removing companies give very similar results.

\begin{figure}
\begin{center}
\includegraphics[width=12cm]{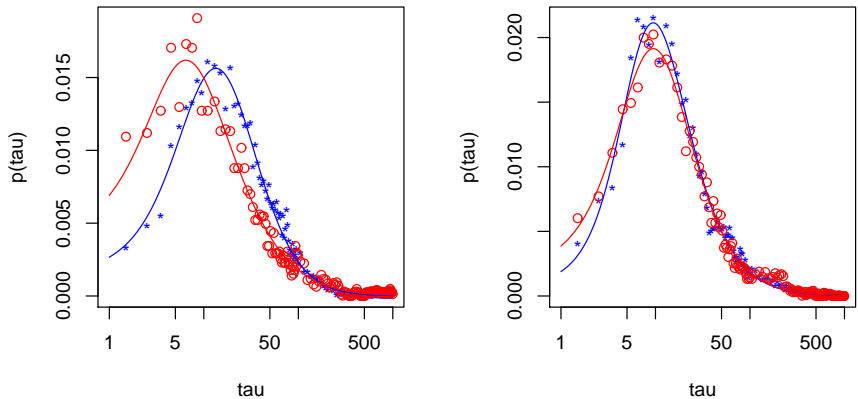}
\end{center}
\caption{Estimated distribution of the first passage time
$\tau_\rho$ for the log price of the artificial index $I$ (left)
and its scrambled version (right). The graphs correspond to $\rho
= +5\%$ (stars) and $\rho = -5\%$ (rings). The solid lines are
fitted generalized gamma density functions.} \label{fig:I_FPT}
\end{figure}

Figure \ref{fig:I_FPT} shows that the index constructed from these
stocks display a gain/loss asymmetry, much like the Dow Jones
index, and that the asymmetry vanishes if we scramble the time
series.

In what follows, we will use this artificial index as a kind of
proxy for a real stock index. This has the advantage that the
individual index constituents are readily available for analysis.
This is unlike the Dow Jones index for which the relative weights
and indeed the set of constituents have changed over time.

\section*{Dependence between constituents during periods of index upturns and downturns}
Here, and in what follows, $\{I_t\}_{t = 0,\ldots,T}$ denotes the
artificial index defined in the previous section.

Inspired by the generalized asymmetric synchronous market model
from Siven, Lins, and Lundbek~Hansen \cite{multiscale}, our
general intuition is that the individual stocks tend to ``move
together'' to a greater degree during index downturns than during
upturns, resulting in more violent downturns than upturns. To
quantify this, we first divide the price history of our artificial
index $I$ into two parts, corresponding to upturns and downturns,
respectively.

Fix a window length $L$ and consider the index return over the
$k$th window,
$$
\Delta I_k = I_{kL} - I_{(k-1)L},
$$
for $k = 1,\ldots,\lfloor T/L\rfloor$, where $\lfloor x \rfloor$
denotes the largest integer smaller than or equal to $x$. We
define the set of indices for which the daily returns belong to a
window over which the index went up,
$$
U = \bigcup_{\{k;\ \Delta I_k > 0\} } \{(k-1)L + 1,\ldots,kL \},
$$
respectively went down,
$$
D = \bigcup_{\{k;\ \Delta I_k < 0\} } \{(k-1)L + 1,\ldots, kL \}.
$$
Note that the sets $U$ and $D$ are disjoint. 

We will consider two measures of dependence between all the
individual stocks and evaluate it for the returns corresponding to
days $t\in U$, and compare that to the same measures evaluated for
days $t\in D$. Before describing the first measure, the mean of
mutual information, we establish some additional notation. Let the
\emph{$n$th index} be defined by
$$
I_{n,t} = \frac{1}{N-1}\sum_{m \neq n} \frac{S_{m,t}}{S_{m,0}}.
$$
The $n$th index is simply the artificial index constructed by
averaging all stocks except the $n$th. Denote the log return at
day $t$ in the $n$th stock and index by $\delta S_{n,t} =
\log(S_{n,t}/S_{n,t-1})$ and $\delta I_{n,t} =
\log(I_{n,t}/I_{n,t-1})$.

\subsubsection*{Mean mutual information}
The mutual information of two discrete stochastic variables $X$
and $Y$ is defined as
$$
M(X,Y) = \sum_{x}\sum_{y} p_{XY}(x,y)
\log\left(\frac{p_{XY}(x,y)}{p_X(x)p_Y(y)} \right),
$$
where $p_{XY}$ denote the joint and $p_X$ and $p_Y$ the marginal
probability functions of $X$ and $Y$. Mutual information can be
written as $M(X,Y) = H(X) + H(Y) - H(X,Y)$, where $H(X)$ and
$H(Y)$ are the marginal entropies, and $H(X,Y)$ is the joint
entropy of $X$ and $Y$, and it is a measure of dependence in the
sense that $X$ and $Y$ are independent if and only if $M(X,Y) =
0$. Mutual information can estimated from a finite set
$\{(X_t,Y_t)\}_{t = 1,\ldots,n}$ of joint samples of $(X,Y)$ in a
number of different ways, see Paninski \cite{mutualInfo}. In the
computations below we apply the most straightforward estimator,
the so-called plug-in estimator.

Let $M_{U,n}$ and $M_{D,n}$ denote the mutual information of the
returns of the $n$th stock and index, estimated from the samples
$\{(\delta S_{n,t}, \delta I_{n,t})\}_{t \in U}$ respectively
$\{(\delta S_{n,t}, \delta I_{n,t})\}_{t \in D}$. We average over
$n$ to obtain the \emph{mean mutual information}, which can be
seen as a measure of the degree of dependence between all the
stocks over periods of upturns, and, respectively, downturns of
the index $I$:
\begin{eqnarray*}
M_U &=& \frac{1}{N}\sum_{n = 1}^N M_{U,n}, \\
M_D &=& \frac{1}{N}\sum_{n = 1}^N M_{D,n}.
\end{eqnarray*}

Figure \ref{fig:MI} shows the mean mutual information for varying
window length --- there is clearly a higher degree of dependence
between the stocks returns during index downturns. However, given
the hypothesis that stocks tend to ``move together'' to a greater
degree during index downturns, with the result that downturns are
more dramatic than upturns, there is a potential problem with the
measure: the mutual information between the $n$th stock and index
is large whenever there is a high degree of dependence, not only
when they tend to move in the same direction. If some stocks tend
to move up when the index moves down, this would moderate the
downturns, contrary to our intuition, and yet result in high
values for the mean mutual information. For this reason, we also
consider a correlation based measure.

\subsubsection*{Mean correlation}
\begin{figure}
\begin{center}
\includegraphics[width=12cm]{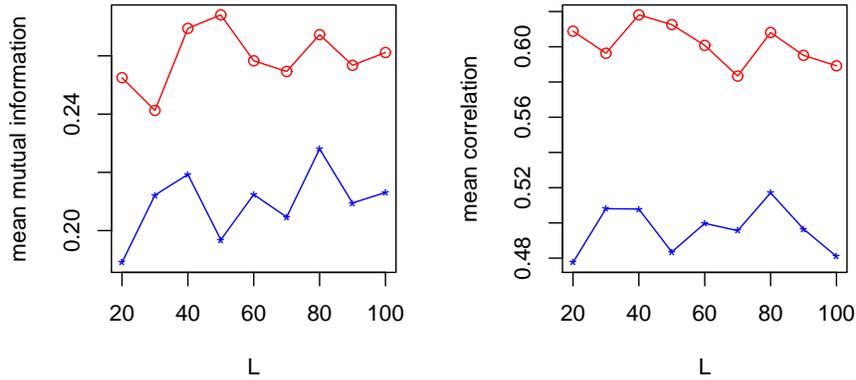}
\end{center}
\caption{The mean mutual information (left) and the mean
correlation (right) for the artificial index, as function of the
window length $L$. The graphs show the mean mutual information and
correlation corresponding to index upturns (stars) respectively
downturns (rings).} \label{fig:MI}
\end{figure}

Let $C_{U,n}$ and $C_{D,n}$ denote the correlation between the
returns of the $n$th stock and index, estimated from the samples
$\{(\delta S_{n,t}, \delta I_{n,t})\}_{t \in U}$ respectively
$\{(\delta S_{n,t}, \delta I_{n,t})\}_{t \in D}$. We average over
$n$ to obtain the \emph{mean correlation}, which can be seen as a
measure of the degree of dependence between all the stocks over
periods of upturns respectively downturns of the index $I$:
\begin{eqnarray*}
C_U &=& \frac{1}{N}\sum_{n = 1}^N C_{U,n}, \\
C_D &=& \frac{1}{N}\sum_{n = 1}^N C_{D,n}.
\end{eqnarray*}
Figure \ref{fig:MI} shows the mean correlation for varying window
length --- this measure of dependence between the stock returns
also show markedly higher values during index downturns that
during index upturns. Contrary to the mean mutual information,
however, the presence of ``defensive'' stocks that move up during
index downturns would give negative contributions.

\section*{Conclusion}
If the gain/loss asymmetry observed for stock indices were a
property of the unconditional distribution of returns, then the
phenomenon should remain invariant under random permutations of
the returns --- this is not the case, as we have demonstrated. We
may begin to rely more confidently on expectations derived from
the generalized asymmetric synchronous market model, which have
previously demonstrated that differences in correlated movements
in index constituents for down-moves and up-moves can give rise to
the kind of temporal dependence structure that produces such
asymmetry. However, there are practical difficulties in exploring
the correlations between the time series of the individual
constituents of real stock indices, since these are not readily
available, so we have shown that the gain/loss asymmetry can also
be reproduced in an artificial stock index constructed as a simple
average of a number of individual stocks.

Considering two different measures of dependence, mean mutual
information and mean correlation, we concluded that there indeed
is a greater degree of dependence between the constituents of the
artificial index during downturns than upturns. This part of our
analysis can be seen as an attempt to overcome some of the general
difficulties in formulating tractable ways of analyzing
non-stationary dependence structure in multivariate stochastic
processes. Future work in the direction of analyzing the dynamics
of the changes in the level of dependence between asset prices
would certainly be interesting --- not least from the perspective
of investors who seek diversification that does not break down at
the worst possible time. For instance, is it possible to design a
localized measure of the level of dependence between stock prices
and zoom in even more on the points in time where it is changing?

\bibliographystyle{plain}

\end{document}